# Adiabatic Invariants in Stellar Dynamics:
# II. Gravitational shocking[1]

Martin D. Weinberg[2]

Department of Physics and Astronomy

University of Massachusetts/Amherst

## ABSTRACT

A new theory of gravitational shocking based on time-dependent perturbation theory shows that the changes in energy and angular momentum due to a slowly varying disturbance are not exponentially small for stellar dynamical systems in general. It predicts significant shock heating by slowly varying perturbations previously thought to be negligible according to the adiabatic criterion. The theory extends the scenarios traditionally computed only with the impulse approximation and is applicable to a wide class of disturbances. The approach is applied specifically to the problem of disk shocking of star clusters.

## 1.   Introduction

A wide variety of astronomical problems require solving for the evolution of a bound stellar system in the external gravitational field of a larger embedding system. Examples include globular clusters on eccentric orbits in the Galaxy, infalling satellite galaxies, galactic star clusters in the disk, and the evolution of young associations in molecular clouds.

In many cases, the characteristic time scale for the external perturbation is neither slower or faster than *all* orbital periods of the bound cluster, but somewhere in between. Slow orbits may be assumed to be static in the time frame of the perturber which leads to the impulse approximation. In the opposite extreme, the perturber is nearly static in the orbit's time frame which is the adiabatic limit. Most studies appeal to the harmonic oscillator model to argue that all orbits with time scales greater than that of the external

---

variation will be invariant; precisely, if the perturbing frequency $\nu$ is smaller than the oscillator frequency $\Omega$ then the net change in energy of a random phase ensemble of oscillators is exponentially small in their ratio, e.g. proportional to $e^{-\Omega/\nu}$. Accordingly, orbits with $\Omega < \nu$ receive a kick by the perturbation computed using the impulse approximation, but the actions of orbits in the adiabatic limit, $\Omega > \nu$, are assumed to be conserved. The seminal study of globular cluster evolution by Ostriker et al.(1972) was based on impulse heating and the adiabatic criterion. This paradigm of a "gravitational shock" is now ubiquitous. Chernoff et al.(1986, see also Chernoff and Shapiro 1987) make the harmonic model explicit, extending the impulse approximation using Spitzer's (1958) treatment of tidal distortions using the linear oscillator model directly.

This paper presents a theory which allows the evolution to be computed for a general time-dependent perturbation regardless of rate. The investigation predicts significant evolution in *both* the adiabatic and impulsive regimes. The explanation for this surprising result is discussed in the preceding companion paper (Paper I) which shows that the harmonic oscillator model does not apply to the general nonlinear multidimensional orbit and describes the failure of adiabatic invariants. To summarize, all stellar systems are multidimensional systems with at least two degrees of freedom. Each independent degree of freedom has a frequency of oscillation, $\Omega_j = \dot{w}_j$. Since any quantity, $Q$, describing a smooth quasi-periodic orbit is well-represented by a rapidly converging Fourier series, $Q = \sum_{\mathbf{l}} Q_{\mathbf{l}} \exp(i\mathbf{l} \cdot \mathbf{w})$, the evolution of an orbit is equivalent to the evolution of a set of pendula whose frequencies are $\mathbf{l} \cdot \mathbf{\Omega}$ where $\mathbf{l}$ is a vector of integers. As long as the frequency of the perturbation remains small compared to $\mathbf{l} \cdot \mathbf{\Omega}$, each pendulum will be adiabatically invariant and the constants of motion (actions) for the original orbit will be conserved. However, if one of those frequencies $\mathbf{l} \cdot \mathbf{\Omega}$ is zero or nearly so, then this term will receive a kick from the perturbation. Since one of the Fourier coefficients will change the original orbit will no longer conserve its actions.

For a definite physical example, take a two-dimensional system which has two distinct frequencies, $\Omega_1$ and $\Omega_2$. If there exists two integers $l_1$ and $l_2$ with $l_1\Omega_1 - l_2\Omega_2 = 0$, then for every $l_2$ periods in degree of freedom 1, the orbit will have executed $l_1$ periods in degree of freedom 2, returning to its original configuration. Since the perturbation frequency is small compared to either $\Omega_1$ and $\Omega_2$, the perturbation "hits" the orbit repetitively at the same phase, cumulatively producing a measurable change in the trajectory. This is similar to a resonance but with arbitrarily small resonant frequency. Now, a realistic stellar system has frequencies continuously distributed in some range. For many orbits, $\mathbf{l} \cdot \mathbf{\Omega}$ will be far from zero for relatively small values of $l_j$ and these orbits will be adiabatically invariant. However, there will almost always be some combinations of small integers for which $\mathbf{l} \cdot \mathbf{\Omega} = 0$ and those orbits will *not* be adiabatically invariant. Averaging over the whole stellar system, the orbits with broken invariants can give an appreciable overall change with magnitude similar to the impulsive contribution.

The large fraction of this paper describes a general method for applying these ideas to stellar systems (§2). We begin with a general statement of the time-dependent perturbation theory for spherical stellar systems and, for a concrete example, apply the scheme to the gravitational shocking of globular clusters by disk passage (§3). Although gravitational shocking is the targeted application, this development holds for any finite-duration perturbation to a spherical stellar system. The approach could be easily extended to disks. The long-term change in energy, action and distribution function are specifically computed. The special cases of binary star evolution and very slow perturbations are discussed (§4). The effects of the new disk shocking formalism presented here on globular cluster evolution is illustrated in Paper III which applies it to Fokker-Planck models.

## 2. Problem statement

Assume a spherical background with the distribution function $f_o(\mathbf{I})$ where $\mathbf{I}$ is the action vector. Let the perturbing potential which causes shocking be given by $V_p(\mathbf{r}, t)$. In action-angle variables, the linearized collisionless Boltzmann equation may be written:

$$\frac{\partial f_1}{\partial t} + \frac{\partial H_o}{\partial \mathbf{I}} \cdot \frac{\partial f_1}{\partial \mathbf{w}} - \frac{\partial f_o}{\partial \mathbf{I}} \cdot \frac{\partial H_1}{\partial \mathbf{w}} = 0 \tag{1}$$

where $H_1 = V_p$ and $\mathbf{w}$ is the vector of angles conjugate to the action vector, and the subscripts '0' and '1' indicate unperturbed and perturbed quantities and $H$ is Hamilton's function.

This equation may be solved for $f_1$ by Fourier transforming equation (1) in angles and Laplace transforming in time. The Fourier transform of a phase space quantity $Q$ is given by:

$$Q = \sum_{\mathbf{l}} Q_{\mathbf{l}}(\mathbf{I}) e^{i \mathbf{l} \cdot \mathbf{w}}, \tag{2}$$

$$Q_1 = \frac{1}{(2\pi)^3} \int d\mathbf{w} e^{-i\mathbf{l} \cdot \mathbf{w}} Q(\mathbf{I}, \mathbf{w}), \tag{3}$$

where $\mathbf{l}$ is a triple vector of integers whose values range over all integers unless otherwise noted. Denote the Laplace transform of a quantity $Q$ by $\hat{Q}$. Now assume that the perturbation vanishes at some time in the past—that is, the cluster used to be far from the perturbation—so that $f_{\mathbf{l}}(t \to -\infty) = 0$. Then we may solve for $\hat{f}_{\mathbf{l}}$:

$$\hat{f}_{\mathbf{l}} = \frac{i\mathbf{l} \cdot \partial f_o / \partial \mathbf{I}}{s + i\mathbf{l} \cdot \mathbf{\Omega}(\mathbf{I})} \hat{V}_p \mathbf{l}. \tag{4}$$

The perturbed distribution function then follows from the inverse Laplace transform, and may be written as

$$f_1 = \sum_{\mathbf{l}} f_{\mathbf{l}}(\mathbf{I}, t) e^{i\mathbf{l}\cdot\mathbf{w}}. \tag{5}$$

Physically, $f_1$ may be thought of as the "wake" in the cluster induced by the perturbation.

Most of the work in solving for $f_1$ is in the step between equation (4) to (5). This calculation does not include the self-consistent change implied by the new distribution function; this could be done (e.g. Weinberg 1989) but only with considerably more computational work. In the next section, we will consider the case of disk shocking and perform the inversion explicitly.

## 3. Disk shocking

Let us consider a globular cluster passing through an one-dimensional slab, representing the disk. More general perturbations may be done similarly; in particular, the time dependent forcing of a cluster on an eccentric orbit will be the subject of a later paper.

Let $v_z$ be the velocity perpendicular to the slab; clearly the transverse velocity is irrelevant. To determine the acceleration relative to the cluster center, the perturbing potential may be expanded about the cluster center of mass to get the tidal strain:

$$V_p \simeq \frac{1}{2} \left.\frac{d^2 V_p}{dz^2}\right|_{z_c(t)} z^2 \tag{6}$$

where $z_c(t)$ is the center of the cluster relative to the slab at time $t$ and $z$ is the vertical position in the cluster relative to its center. The truncation error in equation (6) due to this tidal expansion is $\mathcal{O}([r_{1/2}/h]^2)$. Since $r_{1/2}/h \approx 1/60$, this error is roughly $\sim 0.3\%$.

The development below will treat a Gaussian vertical disk profile,

$$\rho(z) = \rho_0 e^{-(z+v_z t)^2/h^2} \tag{7}$$

where $h$ is the scale height of the slab. The cluster center then crosses the disk's midplane $t = 0$. The Laplace transform (two-sided, e.g. van der Pol & Bremmer 1955) for the Gaussian profile yields

$$\hat{V}_p = \frac{z^2}{2} \frac{V_o}{v_z} \sqrt{\pi} e^{h^2 s^2/4 v_z^2} e^{s z_o/v_z}. \tag{8}$$

The constant quantity $V_o \equiv 4\pi G \rho_o h^2$ follows from Poisson equation (see Appendix for additional details). Other slab profiles may be treated similarly. The exponential slab,

$$\rho(z) = \rho_0 e^{-|z+z_o+v_z t|/h}, \tag{9}$$

will be briefly discussed below for contrast. Although the exponential slab is pathological, having a density cusp at $z = 0$, both models yield similar overall heating rates, remarkably. However, the exponential disk does have different asymptotically behavior than an everywhere smooth model.

To complete the calculation defined in §2, we need the expansion of $z^2$ in action-angle variables. Assume the spatial part of $V_p$ may be expanded in a harmonic series in spherical coordinates: $V_p(r, \theta, \phi) = \sum Y_{nm}(\theta, \phi) f_{nm}(r)$. Using Tremaine & Weinberg (1984 eqn. 54, hereafter TW), each term in the sum may be expanded in action-angle variables:

$$V_p(r, \theta, \phi) \;=\; \sum_{l=0}^{\infty} \sum_{l_1=-\infty}^{\infty} \sum_{l_2=-l}^{l} \sum_{l_3=-l}^{l} V_{l\,l_2\,l_3}(\beta) W_{l\,l_2\,l_3}^{l_1}(\mathbf{I}) e^{i\mathbf{l}\cdot\mathbf{w}} \tag{10}$$

where

$$W_{n\,l\,m}^{k} \;=\; \frac{1}{2\pi} \int_{-\pi}^{\pi} dw_1 \, e^{-ikw_1} f_{nm}(r) e^{il(\psi - w_2)} \tag{11}$$

and $\beta$ is the inclination of the orbital plane (using the notation of TW). The function $f_1(w_1) \equiv \psi - w_2$ is the difference between the mean azimuthal angle $w_2$ and the azimuthal angle in the orbital plane (cf. TW eq. 38). The fact that $m = 0$ for all terms above demands that $l_3 = 0$. The quantity $V_{n\,l\,m} = 0$ for $|l| > n$ or $|m| > n$ and non-zero only when $n + l$ is even. For example, if $n = 0$ then $l = 0$ and the only non-vanishing term is $W_{0\,0\,0}^{l_1}$. Since $z^2 = r^2 \cos^2 \theta$, $f_{nm}(r) = r^2$ which is independent of $n$ and $m$. We may therefore simplify the notation for $W$, defining $X_{l_2}^{l_1} = W_{l\,l_2\,l_3}^{l_1}$ and write

$$z^2 = \sum_{\mathbf{l}=-\infty}^{\infty} \left[ \frac{2}{3} \sqrt{\frac{4\pi}{5}} V_{2\,l_2\,0}(\beta) + \frac{1}{2} \sqrt{4\pi} V_{0\,l_2\,0}(\beta) \right] X_{l_2}^{l_1} e^{i\mathbf{l}\cdot\mathbf{w}}. \tag{12}$$

The inverse Laplace transform of equation (8) may now be straightforwardly performed using equations (4) and (12) and retaining only lowest order contributions to get $f_1(t)$; details are given in the Appendix.

## 4. Phase-averaged perturbations

Most commonly, the long-term ($t \to \infty$) change in a moment or the distribution itself is desired. The former will be explored in this section. The moment changes are expectation values over a specified ensemble. For example, an ensemble of orbits with energy $E$ initially will have a spread of energies after the disk passage; the quantity computed in this section tells us the average change per orbit in the ensemble. On the other hand, the overall evolution is best indicated by the change in the distribution itself. However, equations (4) and (5) show that the phase average of $f_1$ vanishes; the second-order contribution, $f_2$, does not and will be considered in the next section.

Equation (5) and the explicit form of $f_1(t)$ allows us to determine the mean change of any constant of the motion using Hamilton's equations. For any function of phase space, we have (e.g. Goldstein 1950)

$$\frac{dQ}{dt} = \frac{\partial Q}{\partial t} + [Q, H] \tag{13}$$

where [ ] are Poisson brackets. The first term on the right-hand-side vanishes if $Q$ is a constant of the motion. On expanding the previous expression, the lowest order terms vanish by assumption. Retaining only first-order terms yields

$$\frac{dQ}{dt} = \frac{\partial Q}{\partial \mathbf{w}} \cdot \frac{\partial H_1}{\partial \mathbf{I}} - \frac{\partial H_1}{\partial \mathbf{w}} \cdot \frac{\partial Q}{\partial \mathbf{I}}. \tag{14}$$

The angle or phase average of $\dot{Q}$ is then:

$$\left\langle \dot{Q}(\mathbf{I}) \right\rangle = \frac{1}{(2\pi)^3} \int d\mathbf{w} \, \frac{dQ}{dt} f_1 = \sum_{\mathbf{l}} \left[ \frac{\partial Q}{\partial \mathbf{w}} \cdot \frac{\partial V_{p\mathbf{l}}}{\partial \mathbf{I}} - i\mathbf{l} \cdot \frac{\partial Q}{\partial \mathbf{I}} V_{p\mathbf{l}} \right] f_{-\mathbf{l}}(\mathbf{I}, t). \tag{15}$$

In particular, for a component of action we have

$$\left\langle \dot{I}_j(\mathbf{I}) \right\rangle = \sum_{\mathbf{l}} -i l_j V_{p\mathbf{l}} f_{-\mathbf{l}}(\mathbf{I}, t) = \sum_{\mathbf{l}} i l_j V_{p-\mathbf{l}} f_{\mathbf{l}}(\mathbf{I}, t), \tag{16}$$

and since $E = H_o(\mathbf{I})$ we have

$$\left\langle \dot{E}(\mathbf{I}) \right\rangle = \sum_{\mathbf{l}} -i\mathbf{l} \cdot \mathbf{\Omega} V_{p\mathbf{l}} f_{-\mathbf{l}}(\mathbf{I}, t) = \sum_{\mathbf{l}} i\mathbf{l} \cdot \mathbf{\Omega} V_{p-\mathbf{l}} f_{\mathbf{l}}(\mathbf{I}, t). \tag{17}$$

Finally, the phase-averaged change in $Q$ for an ensemble of fixed action $\mathbf{I}$ may then be found by integration: $\langle \Delta Q \rangle = \int_0^\infty dt \, \langle \dot{Q}(\mathbf{I}, t) \rangle$.

### 4.1. Change in energy

As an example, let us evaluate $\langle \Delta E \rangle$ explicitly. For the spherical system discussed here, $I_1$ is the standard radial action, $I_2 = J$ and $I_3 = J_z$ where $J$ and $J_z$ are the total and $z$ projection of the angular momentum. If the system has no net rotation then the distribution function will be independent of $J_z$ and we may integrate over $I_3$ keeping $I_2$ fixed. This motivates changing variables from $I_3$ to inclination $\beta$ defined by $\cos \beta = I_3/I_2$. Then, $d\mathbf{I} = dI_1 dI_2 I_2 d\beta \sin \beta$ and we may define:

$$\langle \langle Q \rangle \rangle = \int_0^\pi d\beta \sin \beta \langle Q \rangle. \tag{18}$$

We now evaluate $\langle\langle\Delta E\rangle\rangle$ resulting from the passage through the slab using the expression for $f_1$ and $V_{p1}(t)$ and find (see Appendix)

$$
\begin{aligned}
\langle\langle\Delta E\rangle\rangle &= -\frac{(2\pi)^3}{4}\frac{V_o^2}{h^4}\sum_1^\infty \delta_{l_3\,0}\left(\frac{1}{15}+\frac{43}{90}\delta_{l_2\,0}\right)|X_{l_2}^{l_1}|^2 \\
&\times\; \mathbf{1}\cdot\frac{\partial f_o}{\partial\mathbf{I}}\frac{\mathbf{l}\cdot\mathbf{\Omega}}{2(v_z/h)^2}\pi e^{-h^2(\mathbf{l}\cdot\mathbf{\Omega})^2/2v_z^2}.
\end{aligned}
\tag{19}
$$

The analogous expression for the exponential disk is also derived in the Appendix:

$$
\begin{aligned}
\langle\langle\Delta E\rangle\rangle &= -\frac{(2\pi)^3}{4}\frac{V_o^2}{h^4}\sum_1^\infty \delta_{l_3\,0}\left(\frac{1}{15}+\frac{43}{90}\delta_{l_2\,0}\right)|X_{l_2}^{l_1}|^2 \\
&\times\; \mathbf{1}\cdot\frac{\partial f_o}{\partial\mathbf{I}}\frac{2\,(\mathbf{l}\cdot\mathbf{\Omega})\,(v_z/h)^2}{[(v_z/h)^2+(\mathbf{l}\cdot\mathbf{\Omega})^2]^2}.
\end{aligned}
\tag{20}
$$

If the distribution function depends on energy alone, $f_o = f(E)$, then

$$
\mathbf{1}\cdot\frac{\partial f_o}{\partial\mathbf{I}}=\mathbf{1}\cdot\frac{\partial E}{\partial\mathbf{I}}\frac{df_o}{dE}=\mathbf{1}\cdot\frac{\partial H}{\partial\mathbf{I}}\frac{df_o}{dE}=\mathbf{1}\cdot\mathbf{\Omega}(\mathbf{I})\frac{df_o}{dE}.
\tag{21}
$$

Upon substituting into the expression for $\langle\langle\Delta E\rangle\rangle$, we find that to lowest order, disk shocking increases the energy for all $E$ if $df_o/dE < 0$.

## 4.2. Comparison of Gaussian and exponential disks

Features of the expressions for $\langle\langle\Delta E\rangle\rangle$ are worth noting. Both equations (19) and (20) scale explicitly as $\langle\langle\Delta E\rangle\rangle \propto [(\mathbf{l}\cdot\mathbf{\Omega})/(v_z/h)]^2$ in the impulsive limit $[(\mathbf{l}\cdot\mathbf{\Omega})/(v_z/h) \ll 1]$ as expected. However in the adiabatic limit, equation (20) is proportional to $[(\mathbf{l}\cdot\mathbf{\Omega})/(v_z/h)]^{-2}$ while equation (19) appears to be exponentially small. However, if there are commensurabilities between frequencies, $\mathbf{l}\cdot\mathbf{\Omega}=0$, the power law behavior does obtain for equation (19) as well for the physical reasons described in Paper I.

To see this, take an ensemble at fixed energy but isotropic in velocity. Integrating the distribution over the ensemble requires an integration over $dJ\,J/\Omega_1(E,J)$. The integrand in this case is proportional to

$$
\frac{(\mathbf{l}\cdot\mathbf{\Omega})^2}{2(v_z/h)^2}e^{-h^2(\mathbf{l}\cdot\mathbf{\Omega})^2/2v_z^2}.
$$

The variation of the exponent depends most strongly on the variation of the ratio of frequencies $\chi \equiv \Omega_2/\Omega_1$ for small $v_z/h$. In these terms, the integral to be done is roughly $\int d\chi\,\chi^3 e^{-\alpha^2(\chi-\chi_o)^2}$ where the weaker dependence of $|X_{l_2}^{l_1}|^2$ on $\chi$ has been ignored. If the variation of $\chi$ includes $\chi_o$, the integral gives a contribution proportional to $\alpha^{-4}$ and leads to the scaling $\langle\langle\Delta E\rangle\rangle \propto [(\mathbf{l}\cdot\mathbf{\Omega})/(v_z/h)]^{-2}$ in the adiabatic limit $[(\mathbf{l}\cdot\mathbf{\Omega})/(v_z/h) \gg 1]$. Explicit numerical evaluation shows that the Gaussian case does have a power law dependence but with steeper slope. Figure 1 compares $\langle\langle\Delta E\rangle\rangle$ for the Gaussian and exponential slabs. Both models yield very similar results suggesting that the response to the gravitational shock is weakly dependent on the disk's vertical structure

Conversely, the reader may be surprised that equation (20) shows no exponential cutoff for incommensurate orbits in the adiabatic regime. This is a consequence of the exponential disk profile, whose force is not smooth at $z = 0$. This discontinuity provides a power at all temporal frequencies, and loosely speaking, creates resonant interactions at all frequencies.

### 4.3. Check by simulation

Figures 2–4 compare the theoretical predictions of equation (19) with restricted n-body simulations for a $W_0 = 5$ King model. The integration accuracy corresponds to an absolute error in energy of approximately $10^{-7}$ per orbit. The parameters roughly describe a typical globular and Galactic disk—$M = 3 \times 10^5\,\mathrm{M_\odot}$, $r_t = 100\,\mathrm{pc}$, $R_g = 8\,\mathrm{kpc}$, $V_c = 200\,\mathrm{km\,s^{-1}}$, and $h = 325\,\mathrm{pc}$. However in order to improve signal/noise, the disk potential is artificially increased by a factor of 3 over its realistic value; this enables the resolution of the highest frequency encounter (Fig. 4). The bin with largest energy is undersampled due to escaping stars and the bins with the smallest energy are affected by the integration accuracy, indicated by the dashed horizontal line. The agreement is good for energies between these two limits.

### 4.4. Application to binary stars

The gravitational shocking theory may be used to compute energy changes for an ensemble of binary stars and is consistent with the standard results obtained for binary star systems (Heggie 1975). In two-body problem, the orbital frequencies are degenerate, $\Omega_1 = \Omega_2$ and $\Omega_3 = 0$, which means that $\mathbf{l}\cdot\mathbf{\Omega}$ vanishes if and only if $\sum l_j$ vanishes. Therefore, the contribution to $\langle\langle\Delta E\rangle\rangle$ will always vanish if $\mathbf{l}\cdot\mathbf{\Omega} = 0$ and the contribution will be exponentially down in the ratio $(\mathbf{l}\cdot\mathbf{\Omega})/(v_z/h)$ as expected for $\mathbf{l}\cdot\mathbf{\Omega} \neq 0$. The energy is a *true* adiabatic invariant for binary stars.

However, the behavior is different for the changes in individual actions $\langle\langle\Delta I_j\rangle\rangle$. Equation (16) gives the change in a single action, $I_j$ and $\langle\langle\Delta I_j\rangle\rangle$ may be derived in the same way as $\langle\langle\Delta E\rangle\rangle$. The distribution function for a single orbit with action $\mathbf{I}_o$ is $f(\mathbf{I}) = \delta(\mathbf{I} - \mathbf{I}_o)/(2\pi)^3$. Integrating by parts and using the Gaussian disk perturbation for definiteness yields:

$$\langle\langle\Delta I_j\rangle\rangle = \frac{1}{4}\frac{V_o^2}{h^4}\frac{\pi}{2(v_z/h)^2}\sum_1^\infty \delta_{l_3\,0}\left(\frac{1}{15} + \frac{43}{90}\delta_{l_2\,0}\right)l_j\,\mathbf{1}\cdot\frac{\partial}{\partial\mathbf{I}}\left[|X_{l_2}^{l_1}|^2 e^{-h^2(\mathbf{1}\cdot\mathbf{\Omega})^2/2v_z^2}\right]. \qquad (22)$$

In the limit $\Omega/(v_z/h) \gg 1$, terms with $\sum l_j \neq 0$ will be exponentially down, but unlike the case for $\Delta E$, the terms with $\sum l_j = 0$ do not vanish. In fact, in this limit the previous equation becomes

$$\langle\langle\Delta I_j\rangle\rangle \approx \frac{1}{4}\frac{V_o^2}{h^4}\frac{\pi}{2(v_z/h)^2}\sum_1^\infty \delta_{l_3\,0}\left(\frac{1}{15} + \frac{43}{90}\delta_{l_2\,0}\right)l_j\,\mathbf{1}\cdot\frac{\partial}{\partial\mathbf{I}}|X_{l_2}^{l_1}|^2. \qquad (23)$$

Evaluating $|X_{l_2}^{l_1}|^2$ explicitly shows that this expression for $\langle\langle\Delta I_j\rangle\rangle$ does not vanish (except for purely radial or circular orbits) and drives orbits to larger eccentricities even though energy is adiabatically conserved ($\langle\langle\Delta I_1\rangle\rangle = \langle\langle\Delta I_r\rangle\rangle \geq 0$ and $\langle\langle\Delta I_2\rangle\rangle = \langle\langle\Delta J\rangle\rangle \leq 0$). It is conceivable that this mechanism modifies the primordial elements of young binaries in the dense environments in which they form and in longer lived star clusters.

### 4.5. Evolution for very slowly changing perturbations

Let us take a general perturbation $V_p(\mathbf{x}, t)$ which is turned on in the past and turned off in the future. Now if the time variation is made *very* slow, e.g. $V_p = V_p(\mathbf{x}, t/\tau)$ for large $\tau$, the scaling for the long-term change in the distribution may be computed directly (originally suggested and computed by Scott Tremaine 1994). The calculation is straightforward and is sketched for $\langle\langle\Delta E\rangle\rangle$ in the Appendix. Integrating over the phase space for the entire model gives the total energy change for the system due to a gravitational shock:

$$\langle\langle\langle\Delta E\rangle\rangle\rangle = -\frac{1}{\tau}4\pi^3\sum_l\sum_1\int_{-\infty}^\infty ds\,s^2\int d\mathbf{I}\,\mathbf{1}\cdot\mathbf{\Omega}\left[\mathbf{1}\cdot\mathbf{\Omega}\frac{\partial f_o}{\partial E}\delta(\mathbf{1}\cdot\mathbf{\Omega}) - l_2\frac{\partial f_o}{\partial J}\delta'(\mathbf{1}\cdot\mathbf{\Omega})\right]|\hat{\Psi}_{l\,l_2\,l_3}^{l_1}|^2. \qquad (24)$$

where $\hat{\Psi}_{l\,l_2\,l_3}^{l_1}(\mathbf{I}, s) = V_{l\,l_2\,l_3}(\beta)W_{l\,l_2\,l_3}^{l_1}(\mathbf{I})$ is the action-angle transform of the perturbation which has been Laplace transformed in time and $\delta$ is the Dirac delta function. Equation (24) shows that the overall change in energy for a cluster or dwarf galaxy scales as $1/\tau$. Comparison with equation (19) and the derivation in the Appendix suggests that $1/\tau$ scaling from the peak amplitude at $\tau \approx 1/\Omega$, should be roughly correct. At any rate, this is very weak adiabatic cutoff, qualitatively and quantitatively different than the expected invariance at large $\tau$.

This scaling shows that no realistic stellar systems are truly invariant to slowly changing external perturbations. For example, the core of a galactic cluster will be modified by the cluster's vertical oscillations even though the oscillation period may be many times the stellar orbital periods. In addition, it is often assumed that a cannibalized dwarf with phase space density higher than its captor will survive disruption. However, this result suggests that gravitational shocking may disrupt even dense dwarf companions.

## 5. Second-order calculation of perturbed distribution function

Here, we will compute the overall change to the distribution function directly which requires extending the calculation to second order. The perturbed distribution function allows the long-term evolution of the galaxy or star cluster to be estimated. This result will be used in the Fokker-Planck implementation in Paper III. The features of the solution are explored below using the singular isothermal sphere.

### 5.1. Derivation

The first-order distribution function was solved in §2. The Boltzmann equation to next order equation is:

$$\frac{\partial f_2}{\partial t} + \frac{\partial H_o}{\partial \mathbf{I}} \cdot \frac{\partial f_2}{\partial \mathbf{w}} + \frac{\partial H_1}{\partial \mathbf{I}} \cdot \frac{\partial f_1}{\partial \mathbf{w}} - \frac{\partial f_1}{\partial \mathbf{I}} \cdot \frac{\partial H_1}{\partial \mathbf{w}} = 0. \tag{25}$$

The $H_2$ term vanishes in the absence of self-gravity, Following §2, the Fourier-Laplace transform of the second order equation becomes

$$s\hat{f}_{2\mathbf{l}} + \mathbf{l} \cdot \mathbf{\Omega} f_1 + \frac{1}{(2\pi)^3} \int d\mathbf{w} e^{-i\mathbf{l} \cdot \mathbf{w}} \int_0^\infty dt e^{-st} \Bigg\{$$

$$\sum_{\mathbf{l}'} \frac{\partial V_{\mathbf{l}'}}{\partial \mathbf{I}} e^{i\mathbf{l}' \cdot \mathbf{w}} \cdot \sum_{\mathbf{l}''} i\mathbf{l}'' f_{\mathbf{l}''} e^{i\mathbf{l}'' \cdot \mathbf{w}} - \sum_{\mathbf{l}''} \frac{\partial f_{\mathbf{l}''}}{\partial \mathbf{I}} e^{i\mathbf{l}'' \cdot \mathbf{w}} \cdot \sum_{\mathbf{l}'} i\mathbf{l}' V_{\mathbf{l}'} e^{i\mathbf{l}' \cdot \mathbf{w}} \Bigg\} = 0 \tag{26}$$

where $\hat{f}_2$ are the expansion coefficients of the second-order distribution function. We may restrict our attention to the $\mathbf{l} = 0$ term which gives the only contribution to the phase-averaged secular change. The integral in $\mathbf{w}$ may then be done immediately, yielding the solution

$$\hat{f}_{2\mathbf{l}=0} = -\frac{1}{s} \int_{-\infty}^\infty dt e^{-st} \sum_{\mathbf{l}'} i\mathbf{l}' \cdot \frac{\partial}{\partial \mathbf{I}} \left( V_{-\mathbf{l}'} f_{\mathbf{l}'} \right). \tag{27}$$

Using the action-angle expansion of the Gaussian disk perturbation, the time integral may be performed explicitly.

We must now do the inverse Laplace transform to get from $\hat{f}_{2\,\mathbf{l}=0}(\mathbf{I}, s)$ to $f_{2\,\mathbf{l}=0}(\mathbf{I}, t)$. The joint domain of convergence is $\Re(s) > 0$ and we deform the contour to the imaginary axis to perform the inverse transform for $t \to \infty$. Since there is a pole along the imaginary axis, we may divide the contribution in to the principal part and residue. Changing variables to $s = iy$ shows that the principal part does indeed exist. However, it has a factor of the form $\lim_{t\to\infty} \exp(iyt)$. Because the rest of the integrand is analytic on physical grounds, the principle part vanishes. The final expression consisting of the residue contribution alone is:

$$f_{w\,\mathbf{l}=0} = \frac{V_o^2}{4h^4} \frac{\pi}{2(v_z/h)^2} \sum_{\mathbf{l}=-\infty}^{\infty} \mathbf{l} \cdot \frac{\partial}{\partial \mathbf{I}} \left\{ e^{-(\mathbf{l}\cdot\boldsymbol{\Omega})^2/2(v_z/h)^2} \mathbf{l} \cdot \frac{\partial f_o}{\partial \mathbf{I}} [z^2]_{\mathbf{l}} [z^2]_{-\mathbf{l}} \right\} \tag{28}$$

having used the symmetry in $\mathbf{l} \cdot \boldsymbol{\Omega}$ over the summation in $\mathbf{l}$. The quantity $[z^2]_{\mathbf{l}}$ is the $\mathbf{l}$ Fourier component of the action-angle expansion of $z^2$ whose expansion is given in §3. Integrating over all orbital planes ($\beta$), this becomes:

$$\begin{aligned}
\langle f_{w\,\mathbf{l}=0} \rangle &= \frac{V_o^2}{4h^4} \frac{\pi}{2(v_z/h)^2} \sum_{\mathbf{l}=-\infty}^{\infty} \delta_{l_3\,0} \left( \frac{1}{15} + \frac{43}{90}\delta_{l_2\,0} \right) \mathbf{l} \cdot \frac{\partial}{\partial \mathbf{I}} \Bigg\{ \\
&\qquad e^{-(\mathbf{l}\cdot\boldsymbol{\Omega})^2/2(v_z/h)^2} \mathbf{l} \cdot \frac{\partial f_o}{\partial \mathbf{I}} \left| X_{l_2}^{l_1} \right|^2 . \Bigg\}
\end{aligned} \tag{29}$$

Finally, we may write the distribution function in terms of $E$ and $J$ since this is the most common form with

$$M = (2\pi)^3 \int dE \int dJ\, J \frac{1}{\Omega_1(E, J)} f(E, J).$$

With these conventions, equation (29) becomes:

$$\begin{aligned}
\langle f_2 \rangle &= \frac{V_o^2}{4h^4} \frac{\pi}{2(v_z/h)^2} \sum_{\mathbf{l}=-\infty}^{\infty} \delta_{l_3\,0} \left( \frac{1}{6} + \frac{7}{36}\delta_{l_2\,0} \right) \left( \mathbf{l} \cdot \boldsymbol{\Omega} \frac{\partial}{\partial E} + l_2 \frac{\partial}{\partial J} \right) \Bigg\{ \\
&\qquad e^{-(\mathbf{l}\cdot\boldsymbol{\Omega})^2/2(v_z/h)^2} \left( \mathbf{l} \cdot \boldsymbol{\Omega} \frac{df_o}{dE} + l_2 \frac{\partial}{\partial J} \right) \left| X_{l_2}^{l_1} \right|^2 \Bigg\}.
\end{aligned} \tag{30}$$

In practical applications, $\langle f_2 \rangle$ must be evaluated numerically. Most of the time in evaluating $\langle f_2 \rangle$ and $\langle\langle \Delta E \rangle\rangle$ is the computation of $|X_{l_2}^{l_1}|^2$. An easily implemented numerical approach is described in the Appendix.

## 5.2.   Isothermal model

The singular isothermal sphere is similar to the King model over part of its energy range but because it is scale free, the disk-shocking calculation need only be done for a single energy over the range of desired $\nu \equiv v_z/h$. The simplicity of this model makes it a good tool for exploring the basic features of the theory.

The singular isothermal sphere may be defined as follows. If one takes the potential to be

$$U(r) = 2\sigma^2 \ln r \tag{31}$$

then from Poisson's equation

$$\rho(r) = \frac{\sigma^2}{2\pi G r^2}, \tag{32}$$

$$M(r) = \frac{2\sigma^2}{G} r, \tag{33}$$

and the distribution function is

$$f(E) = \frac{2\sigma^2}{(4\pi)^3 \sqrt{\pi}} e^{-E/\sigma^2} \tag{34}$$

normalized consistently with $\rho(r)$. However, we will take $f(E) = \exp(-E/\sigma^2)$ for convenience in the calculations below.

Using these relations, we may define scale-free quantities. Defining $\kappa \equiv J/J_{max}(E)$, where $J_{max}(E)$ is the maximum orbital angular momentum at fixed energy $E$, it is easy to show that

$$J_{max}(E) = J_{max}(0) e^{E/2\sigma^2},$$
$$r(E, \kappa) = r(0, \kappa) e^{E/2\sigma^2},$$

and

$$\Omega_j(E, \kappa) = \Omega_j(0, \kappa) e^{-E/2\sigma^2}.$$

The quantity $r$ is the radius of an orbit with $E$ and $\kappa$ at a particular phase, e.g. a turning point. Substituting these expressions into equation (30) shows that $\langle f_2 \rangle$ is independent of energy except through the factor $\exp[-(\mathbf{l} \cdot \mathbf{\Omega})^2/2\nu^2]$. Therefore, $\langle f_2 \rangle$ as a function of $E$ for $\nu = 1$ is equivalent to $\langle f_2 \rangle$ evaluated at $E = 0$ for $\nu = \exp(E/2\sigma^2)$. This is also true for $\langle\langle \Delta E \rangle\rangle$. Other quantities may be scaled similarly.

Figure 5 shows the change in energy in an ensemble where orbits have single value of $E$ initially, $\langle\langle \Delta E \rangle\rangle$. Note that this is *not* the same as the change in energy of the ensemble with post-shock energy $E$. The top panel is multiplied by $\nu^2$ to show the asymptotic behavior in the impulsive regime, $\langle\langle\Delta E\rangle\rangle \propto \nu^{-2}$. The lower panel is scaled to show $\langle\langle\Delta E\rangle\rangle$ at $E = 0$ as a function of $\ln \nu^2$. Detailed analyses show that $\langle\langle\Delta E\rangle\rangle \propto \nu^{3\cdot}$ in the adiabatic regime ($E \lesssim -1$).

The lower panel also indicates significant change in energy per orbit inside of the adiabatic boundary. The features in this figure are due to individual contributions ($l_1, l_2$). The individual contributions are shown separately in Figure 6. The terms $(1, 0)$ and $(0, 2)$ contribute the peak on the right hand side and comprise most of the impulsive contribution. These components have incommensurate frequencies and similar profiles with the lowest order having the highest amplitude. The peak on the left hand side is dominated by the $(1, -2)$ term and is largely in the adiabatic regime; this contribution is due to the accidental degeneracy as discussed in Paper I. The $(2, -2)$ term is also in this regime but is of lesser importance. Note the different profiles for each of these and the incommensurate terms.

Figure 7 shows $\nu^2\langle f_2 \rangle$ as a function of $E/\sigma^2$ for $\nu = 1$ and $\langle f_2 \rangle$ at $E = 0$ as a function $\ln \nu^2$. The value of $\Omega$ is approximately 2.5 for $E = 0$ which corresponds to $E = \ln \nu^2 \approx 2$ at the boundary between the adiabatic and impulsive regimes, $\Omega/\nu \approx 1$. Again, the asymptotic impulsive behavior $\langle f_2 \rangle \propto \nu^{-2}$ is obtained for $E \gtrsim 2$. The overall peak contribution is in the transition between the adiabatic and impulsive regimes. Contributions of individual terms ($l_1, l_2$) dominating $\langle f_2 \rangle$ is shown in Figure 8. The incommensurate terms $(0, 2)$ and $(1, 0)$ contribute in the impulsive regime; they are effectively zero for $E \lesssim 0$ and decrease quickly inside of $E \approx 2$. The commensurate $(1, -2)$ and $(2, -2)$ terms contribute at smaller binding energies. The $(1, -2)$ term is the strongest of these and contributes positive density. Responses with positive pattern speed [all but $(1, -2)$] shift stars outward to higher energy and vice versa. Clearly the relative amplitudes of individual contributions depends on the particular distribution and therefore the details of the isothermal model are suggestive only. As a function of passage frequency, the lower panel in Figure 7 shows that $\langle f_2 \rangle$ increases (decreases) in the adiabatic (impulsive) regime.

Since the singular isothermal sphere has infinite extent, it is not meaningful to compute the total heating (because of the semi-infinite interval in energy space in the impulsive regime). For a truncated cluster, the net effect of the shock on the cluster depends on the truncation energy relative to the peak in energy transfer. We may use the singular isothermal sphere as a guide to the possible effects of a gravitational shock. First, if peak feature in Figures 5 and 7 occurs at relatively large binding energies, the overall contribution will be dominated by the impulsive interactions. However, if the feature occurs in the halo (e.g. small $\nu$), heating may be dominated by adiabatic interactions. Secondly, although the interaction increases the energy of an initially monoenergetic ensemble (cf. Fig. 5), the r.m.s. change is large compared to the net change. In other words, individual particles are both gain and lose energy. If the peak occurs at sufficiently low energy in the halo, the heated orbits may escape leaving a cluster with overall higher binding energy. It is therefore possible to "cool" the cluster in a disk shock.

## 6.  Thick-disk and bulge shocking

Thick-disk shocking is a straightforward variant and somewhat easier to compute than the calculation above. Since the perturbation has the form $V_p = g(t)z^2$, if the cluster is dynamically part of the disk then $g(t)$ will be a periodic function of time and may be expanded as a Fourier series in its vertical oscillation period, $P$:

$$g(t) = \sum_{k=-\infty}^{\infty} g_k e^{ik\omega t}, \tag{35}$$

where $\omega = 2\pi/P$. The Laplace transform of a Fourier series is trivial and in most cases, $g_k$ will converge rapidly with increasing $|k|$.

The calculation is analogous for bulge shocking but the potential perturbation expansion will be more general than the $z^2$ dependence and include all second order moments ($Y_{2m}$ terms). In addition the Fourier expansion of $g(t)$ will have two indices corresponding to the radial and azimuthal motions of the cluster orbit. This approach *can* easily include the centrifugal potential but not the velocity-dependent Coriolis force.

External heating will most likely play a different role in evolving clusters which are kinematically a halo component or a disk or thick-disk component. Eccentricity is likely to critical also, as Aguilar et al.(1988) have pointed out for bulge shocking. A detailed investigation of gravitational shocking for a globular cluster on a general orbit—implementing disk, orbit and bulge shocking—is in progress.

## 7.  Summary

Paper I showed that adiabatic invariants are NOT exponentially controlled[3] for all orbits if the number degrees of freedom for the system is greater than one. Consequently for a stellar system, some orbits will be strongly perturbed even if the characteristic frequencies $\Omega$ are much larger than the perturbing frequency $\nu$. This leads to measurable

---

[3]proportional to $e^{-\omega/\nu}$ where $\omega$ and $\nu$ are the characteristic system perturbation frequencies

overall heating of a star cluster or galaxy in the adiabatic regime. The change in energy of an initially monoenergetic ensemble scales approximately as $(\nu/\Omega)^\beta$ $(2 \lesssim \beta \lesssim 3$ depending on the model) compared to $(\nu/\Omega)^{-2}$ for the impulsive regime; the two connect smoothly at $\nu \approx \Omega$. For a perturbation which changes slowly over time $\tau$, the total change in kinetic energy due to the gravitational shock scales as $1/\tau$; there is no sharp adiabatic cutoff.

Heating in the adiabatic regime requires frequencies that are irrational ratios of each other except on discrete surfaces in phase space.[4] This is a very weak condition, true for nearly all commonly used cluster and galaxy models. A binary star system is counter example since $\Omega_1 = \Omega_2 \neq 0, \Omega_3 = 0$ for all energies. Nevertheless, even though the energy of a binary is adiabatically invariant, the method shows that individual actions for a binary system are not invariants. For example, a slow perturbation *can* change a binary's eccentricity in the adiabatic regime.

In addition to energy changes, we derived the long-term change to the distribution itself which requires extending the perturbation theory to second order. The expression for the shocked distribution is easily incorporated into a Fokker-Planck simulation, and this is the subject of Paper III.

Since the theory predicts gravitational shocking over a much wider range of encounter rates and at larger magnitude than previous estimates, a variety of scenarios in addition to standard disk shocking may require revision. First, the adiabatic criterion does not abruptly limit the heating of clusters due to the relatively slow vertical motion in the disk. The development may be modified to account for a periodic perturbation appropriate to shocking a thick-disk globular cluster and by cumulative encounters with GMCs. Secondly, the time-dependent external force of a globular on an eccentric orbit is also a gravitational shock, often called a "bulge shock" in its extreme form. This has already been discussed by Aguilar et al.(1988) among others but the current work suggests that its importance may be even greater. The same effects may be important for dwarf galaxies on moderately eccentric orbits and may be easily applied to disks. This approach will not account for the Coriolis force but will be correct for a radial cluster orbit and detailed simulations (Murali et al.1994) suggest the Coriolis force causes no dominant effects. The seminal shocking problem, the response of a star cluster to a passing molecular clouds, a point-mass fly by (Spitzer 1958) may be performed similarly to include the additional heating. The development for binary star evolution is sketched in §4.4 and may be useful for computing the effects of protostellar evolution.

---

[4]commensurabilities $\mathbf{l} \cdot \mathbf{\Omega} = 0$ where $\mathbf{l}$ a vector of integers

I thank David Chernoff, Greg Fahlman, Chigurupati Murali, Doug Richstone and Scott Tremaine for stimulating discussions, and the Institute for Theoretical Physics in Santa Barbara for its hospitality. This work was supported in part by NSF grant PHY89-04035 to ITP and NASA grant NAGW-2224.

## A.  Gaussian slab profile

The potential for the Gaussian slab, $\rho(z) = \rho_o \exp(-z^2/h^2)$, is

$$V = \frac{V_o}{2}\left[e^{-z^2/h^2} - 1 + \sqrt{\pi}\frac{z}{h}\text{erf}\left(\frac{z}{h}\right)\right] \tag{A1}$$

where $V_o \equiv 4\pi G\rho_o h^2$. Using the tidal prescription in §3 with $z = z_c(t) = v_z t$, the (two-sided) Laplace transform is

$$\hat{V}_p = \frac{z^2}{2}\frac{V_o}{v_z}\sqrt{\pi}e^{h^2 s^2/4v_z^2}. \tag{A2}$$

The inverse transform required to derive the distribution function (cf. eq. 5) may be performed by deforming the contour to the imaginary axis, since the joint domain of convergence is $\Re(s) > 0$. Simplifying, the required integral takes the form

$$Z(t) = \frac{1}{2\pi}\int_{-\infty}^{\infty} dy\, \frac{1}{i(y + \mathbf{l}\cdot\mathbf{\Omega})}e^{-h^2 y^2/4v_z^2}e^{-iyt}. \tag{A3}$$

We now separate the integral to the residue and principle part denoted by $\mathcal{R}$ and $\mathcal{P}$ respectively. The residue trivial. The principle part may be computed by defining $z \equiv y + \mathbf{l}\cdot\mathbf{\Omega}$, splitting the integrand into the two semi-infinite intervals $(-\infty, 0]$ and $[0, \infty)$, changing $z \to -z$ in the former term and combining. Putting these together with equation (4) yields

$$f_1 = \frac{V_o}{2h^2}\frac{\sqrt{\pi}}{v_z/h}i\mathbf{l}\cdot\frac{\partial f_o}{\partial\mathbf{I}}\left[z^2\right]_1 e^{-h^2(\mathbf{l}\cdot\mathbf{\Omega})^2/4v_z^2}e^{i(\mathbf{l}\cdot\mathbf{\Omega})t}\left\{\frac{1}{2} + \right.$$
$$\left.\frac{1}{\pi i}\int_0^\infty \frac{dz}{z}e^{-h^2 z^2/4v_z^2}\left[\sinh\left(\frac{\mathbf{l}\cdot\mathbf{\Omega}}{2v_z^2/h^2}\right)\cos tz - i\cosh\left(\frac{\mathbf{l}\cdot\mathbf{\Omega}}{2v_z^2/h^2}\right)\sin tz\right]\right\}. \tag{A4}$$

The expression $[z^2]_1$ denotes the $\mathbf{l}$ Fourier coefficient in the action-angle expansion of $z^2$ (eq. 12).

We now evaluate $\langle\langle\Delta E\rangle\rangle$ for the using the expression for $f_{\mathbf{l}}$ and $V_{p\mathbf{l}}(t)$. In particular, we need to do the integral

$$Y \equiv \int_{-\infty}^{\infty} dt\, e^{h^2(t+z_o/v_z)^2/v_z^2}(\mathcal{R} + \mathcal{P}). \tag{A5}$$

Performing the algebra, one finds

$$Y \equiv e^{-h^2(\mathbf{l}\cdot\mathbf{\Omega})^2/2v_z^2} \frac{\sqrt{\pi}}{v_z/h} \left\{ 1 + \frac{1}{\pi i} \int_0^\infty \frac{dz}{z} e^{-h^2 z^2/2v_z^2} \sinh\left(\frac{\mathbf{l}\cdot\mathbf{\Omega}}{v_z^2/h^2}\right) \right\}. \quad (A6)$$

From its definition in TW, we have $V_{n l_2 l_3}(\beta) = r_{l_2 l_3}^n(\beta) Y_{n l_2}(\pi/2, 0) i^{l_3 - l_2}$. Using the orthogonality of the rotation matrices, $\int_0^\pi d\beta \sin\beta r_{l_2 l_3}^n(\beta) r_{l_2 l_3}^{n'}(\beta) = 2/(2l+1)\delta_{n n'}$, and the relations $V_{n-l-m} = (-1)^m V_{n l m}$ and $W_{n l m}^{k*} = (-1)^m W_{n-l-m}^{-k}$ we see that only terms with the same value $n$ will contribute. As in §3, only the terms even in $\mathbf{l}\cdot\mathbf{\Omega}$ contribute to $\langle\langle\Delta E\rangle\rangle$, which eliminates the second term in $Y$. So finally, the expression becomes (eq. 19):

$$\langle\langle\Delta E\rangle\rangle = -\frac{(2\pi)^3}{4} \frac{V_o^2}{h^4} \sum_1^\infty \delta_{l_3 0} \left(\frac{1}{15} + \frac{43}{90}\delta_{l_2 0}\right) |X_{l_2}^{l_1}|^2$$
$$\times \mathbf{l} \cdot \frac{\partial f_o}{\partial \mathbf{I}} \frac{\mathbf{l}\cdot\mathbf{\Omega}}{2(v_z/h)^2} \pi e^{-h^2(\mathbf{l}\cdot\mathbf{\Omega})^2/2v_z^2}. \quad (A7)$$

To minimize the computational expense, the sum over $\mathbf{l}$ may be written as

$$\sum_1^\infty = 2 \sum_{\substack{l_1=0 \\ l_2=-2,0,2 \\ l_3=-\infty}}^{\infty}{}' \quad .$$

The restriction in the limits of the sum follow from the properties of $V_{n l m}$ and the prime is to remind that $l_2 \neq -2$ if $l_1 = 0$.

## B. Exponential slab profile

The potential of the disk profile,

$$\rho(z) = \rho_0 e^{-|z + v_z t|/h},$$

follows from direct integration of Poisson's equation yielding

$$V(z) = 4\pi\rho_0 h^2 \left[\frac{|z|}{h} + e^{-|z|/h} - 1\right]. \quad (B8)$$

Note that the third derivative of the potential or second of the force is discontinuous (see text of a discussion of the implications).

The Laplace transform is

$$\hat{V}_p = \frac{z^2}{2} \frac{V_o}{h^2} \left[\frac{1}{v_z/h - s} + \frac{1}{v_z/h + s}\right] \quad (B9)$$

with the domain of converge $-v_z/h < \Re(s) < v_z/h$. Using equation (4), we now take the inverse Laplace transform to get $f_1(t)$. Since the domain of convergence for $f_1$ is $\Re(s) > 0$, the joint domain is $0 < \Re(s) < v_z/h$ and since the poles are explicit, the inverse is simple. To facilitate evaluation, the contour may be deformed to $c \to \infty$ if $t < t_o$ and $c \to -\infty$ if $t > t_o$. One finds

$$f_1(t \leq t_o) = i\mathbf{l} \cdot \frac{\partial f_o}{\partial \mathbf{I}} \frac{V_o}{2h^2}[z^2]_1 \left[ \frac{e^{v_z t/h}}{v_z/h + i\mathbf{l} \cdot \mathbf{\Omega}} \right] \tag{B10}$$

$$f_1(t > t_o) = i\mathbf{l} \cdot \frac{\partial f_o}{\partial \mathbf{I}} \frac{V_o}{2h^2}[z^2]_1 \left[ \frac{2v_z/h e^{-i\mathbf{l} \cdot \mathbf{\Omega} t}}{\Delta^2} - \frac{e^{-v_z t/h}}{v_z/h - i\mathbf{l} \cdot \mathbf{\Omega}} \right], \tag{B11}$$

where $\Delta^2 \equiv (v_z/h)^2 + (\mathbf{l} \cdot \mathbf{\Omega})^2$.

Proceeding as in the previous section, the time integration may be done by breaking the interval into the two parts $(-\infty, 0]$ and $[0, \infty)$. Again since only terms even in $\mathbf{l}$ contribute by symmetry, only one term contributes. Integrating over $\beta$ then yields (eq. 20)

$$\langle\langle \Delta E \rangle\rangle = -\frac{(2\pi)^3}{4} \frac{V_o^2}{h^4} \sum_1^\infty \delta_{l_3 0} \left( \frac{1}{15} + \frac{43}{90} \delta_{l_2 0} \right) |X_{l_2}^{l_1}|^2$$

$$\times \quad \mathbf{l} \cdot \frac{\partial f_o}{\partial \mathbf{I}} \frac{2 (\mathbf{l} \cdot \mathbf{\Omega}) (v_z/h)^2}{[(v_z/h)^2 + (\mathbf{l} \cdot \mathbf{\Omega})^2]^2}. \tag{B12}$$

## C.   $\langle\langle \Delta E \rangle\rangle$ for very slow perturbations

In order to show $\langle\langle \Delta E \rangle\rangle$ is proportional to $1/\tau$ as $\tau \to \infty$, the same steps in the previous two appendices are repeated formally, expanding at the end in powers of $t/\tau$. In particular to start, we need to do the integral

$$Z \equiv \int_{-\infty}^{\infty} V_{p-1}(t) f_1(t), \tag{C13}$$

where $f_1(t)$ is given by the inverse transform of equation (4) and the perturbed potential is $V_p(\mathbf{x}, t) = U(\mathbf{x}, t/\tau)$. Assuming that the Laplace transform for $V_p$ converges at least for $Re(s) > 0$, substituting the explicit expressions for inverse transforms into equation (C13) and using Cauchy's theorem gives

$$Z = \int_{-\infty}^{\infty} dt\, U(\mathbf{x}, t/\tau) \int_{-\infty}^{\infty} dt'\, e^{-i\mathbf{l} \cdot \mathbf{\Omega}(t-t')} \Theta(t-t') U(\mathbf{x}, t'/\tau)$$

$$= \int_0^{\infty} dq\, e^{-i\mathbf{l} \cdot \mathbf{\Omega} q} \int_{-\infty}^{\infty} dt\, U(\mathbf{x}, t/\tau) U(\mathbf{x}, (t+q)/\tau) \tag{C14}$$

where $\Theta$ denotes the Heavyside function and the second equality follows from the change of variables $q = t - t'$ followed by $t \to t + q$. Finally, using the convolution theorem and the even symmetry of the result in $\mathbf{l}$ gives

$$Z = \frac{\tau^2}{2} |\tilde{U}(\mathbf{x}, \mathbf{l} \cdot \boldsymbol{\Omega} \, \tau)|^2. \tag{C15}$$

Now consider the expression

$$\int d\omega \, \omega^2 |\tilde{U}(\mathbf{x}, \omega \tau)|^2 f(\omega)$$

for some arbitrary function $f(\omega)$. In the limit $\tau \to \infty$, the quantity $|\tilde{U}(\mathbf{x}, \omega)|^2$ represents a positive definite distribution in $\omega$ peaked at $\omega = 0$; this may be represented as a Fourier transform in $\omega$. Using this explicitly and after some manipulation, one finds

$$\int d\omega \, \omega^2 |\tilde{U}(\mathbf{x}, \omega)|^2 f(\omega) = \frac{1}{\tau} f(0) \int ds \, s^2 |\tilde{U}(\mathbf{x}, s)|^2 \tag{C16}$$

and therefore in the large $\tau$ limit

$$\omega^2 |\tilde{U}(\mathbf{x}, \omega)|^2 = \frac{1}{\tau} \delta(\omega) \int ds \, s^2 |\tilde{U}(\mathbf{x}, s)|^2. \tag{C17}$$

Similarly, one also finds

$$\omega |\tilde{U}(\mathbf{x}, \omega)|^2 = -\frac{1}{\tau} \delta'(\omega) \int ds \, s^2 |\tilde{U}(\mathbf{x}, s)|^2. \tag{C18}$$

Putting this together gives

$$\langle\langle \Delta E \rangle\rangle = -\frac{1}{2} \frac{1}{\tau} \sum_{\mathbf{l}} \mathbf{l} \cdot \boldsymbol{\Omega} \left[ \mathbf{l} \cdot \boldsymbol{\Omega} \frac{\partial f_o}{\partial E} \delta(\mathbf{l} \cdot \boldsymbol{\Omega}) - l_2 \frac{\partial f_o}{\partial J} \delta'(\mathbf{l} \cdot \boldsymbol{\Omega}) \right] \int ds \, s^2 |\tilde{U}_{\mathbf{l}}(\mathbf{I}, s)|^2. \tag{C19}$$

## D. Computational considerations

Calculation of the central equations of this paper, equation (30) in particular, should be straightforward to anyone familiar with solving for cluster evolution using the Fokker-Planck approach. The only new detail is the computation of the potential transform $X_{l_2}^{l_1}$. The following procedure solves for all needed quantities by direct quadrature and a set of coupled ordinary differential equations.

1. Given $E$ and $J$, determine the turning points $r_a$ and $r_p$ by solving the standard implicit equation $0 = 2[E - U(r)] - J^2/r^2$.

Table 1: System of equations for computing potential transforms

| Variable | Equation | Initial condition |
|----------|----------|-------------------|
| $r$ | $\dot{r} = y$ | $r(t=0) = r_p$ |
| $y = \dot{r}$ | $\dot{y} = -\dfrac{dU}{dr} + \dfrac{J^2}{r^3}$ | $y(t=0) = 0$ |
| $f$ | $\dot{f} = \dfrac{J}{r^2} - \Omega_2$ | $f(t=0) = 0$ |
| $w_1$ | $\dot{w}_1 = \Omega_1$ | $w_1(t=0) = 0$ |
| $X_{l_2}^{l_1}$ | $\dot{X}_{l_2}^{l_1} = \dfrac{\Omega_1}{\pi}\cos(l_1 w - l_2 f) r^2$ | $X_{l_2}^{l_1}(t=0) = 0$ |

2. Determine the orbital frequencies by direct quadrature:

$$
\begin{aligned}
\frac{\pi}{\Omega_1} &= \int_{r_p}^{r_a} dr \frac{1}{\sqrt{2[E - U(r)] - J^2/r^2}}, \\
\Omega_2 &= \frac{\Omega_1}{\pi} \int_{r_p}^{r_a} dr \frac{1}{\sqrt{2[E - U(r)] - J^2/r^2}} \frac{J}{r^2},
\end{aligned}
$$

(D20)

3. The potential transform may now be done by simultaneously integrating the radial component equations of motion for the orbit, the differential expressions for the radial angle, the difference between the true and mean azimuthal angles, and the potential transform itself. The set is shown in Table 1. The integration is assumed to begin at apocenter at $t = 0$. For a particular pair $(l_1, l_2)$, the system has five coupled equations. However, since only a small number of terms $l_1, l_2$ contribute as shown in the previous section and §5, all $n$ of these may be done simultaneously, giving a set of $4 + n$ equations.

# REFERENCES

Aguilar, L., Hut, P., and Ostriker, J. P. 1988, Astrophys. J., 335, 720.

Chernoff, D. F., Kochanek, C. S., and Shapiro, S. L. 1986, Astrophys. J., 309, 183.

Chernoff, D. F. and Shapiro, S. L. 1987, Astrophys. J., 322(1, part 1), 113.

Goldstein, H. 1950, *Classical Mechanics*, Addison-Wesley, Reading, Massachusetts.

Heggie, D. C. 1975, Mon. Not. R. Astron. Soc., 173, 729.

Murali, C., Chernoff, D. F., and Weinberg, M. D. 1994, *Self-consistent evolution of star clusters in the Galactic tidal field*, in preparation.

Ostriker, J. P., Spitzer, L., and Chavalier, R. A. 1972, Astrophys. J., Lett., 176, L47.

Spitzer, L. 1958, Astrophys. J., 127, 17.

Tremaine, S. 1994, *private communication*, unpublished.

Tremaine, S. and Weinberg, M. D. 1984, Mon. Not. R. Astron. Soc., 209, 729.

van der Pol, B. and Bremmer, H. 1955, *Operational Calculus*, Cambridge University Press, second edition edition.

Weinberg, M. D. 1989, Mon. Not. R. Astron. Soc., 239, 549.

Weinberg, M. D. 1994a, *Adiabatic Invariants in Stellar Dynamics: I. Basic concepts*, submitted (Paper I).

Weinberg, M. D. 1994b, *Adiabatic Invariants in Stellar Dynamics: III. Application to globular cluster evolution*, submitted (Paper III).

## FIGURE CAPTIONS

Fig. 1.— Comparison of $\langle\langle \Delta E \rangle\rangle$ for a $W_0 = 5$ King model perturbed by a Gaussian slab (solid) and exponential slab (dashed) for $v_z/h = 0.03, 1.0, 10.0$ (labeled at right). The amplitude is arbitrary.

Fig. 2.— Comparison of direct simulation to perturbation theory calculation. The histogram shows the results of $\langle\langle \Delta E \rangle\rangle$ computed from a direct integration in a fixed potential with 200K particles realized from a $W_0 = 5$ King model. The dotted line shows the mean integration error per orbit in each bin. The solid curve shows the predicted relation using the formula in the notes.

Fig. 3.— Same as Fig. 2 but with the passage frequency decreased by a factor of 3.

Fig. 4.— Same as Fig. 2 but with the passage frequency increased by a factor of 10.

Fig. 5.— The energy change for ensembles whose orbits have the same value $E$ initially. The top panel shows $\langle\langle \Delta E \rangle\rangle$ multiplied by $\nu^2$ but for true $E$ and the bottom panel scaled to $E = 0$.

Fig. 6.— As in the lower panel of Fig. 7 but but individual contributions $(l_1, l_2)$ are shown. The total is omitted for clarity.

Fig. 7.— The second-order perturbed distribution function multiplied by $\nu^2 = (v_z/h)^2$ (top panel) and scaled to $E = 0$ as a function of $\ln \nu^2$ (lower panel). The lower scale may used to read $\nu^2 \langle f_2 \rangle$ as a function of $E$ or as a function of $\nu$ at $E = 0$.

Fig. 8.— As in the upper panel of Fig. 7 but individual contributions $(l_1, l_2)$ are shown along with the total. Other terms are negligible.